\documentclass[journal]{IEEEtran}
\usepackage[utf8]{inputenc}
\usepackage{cite}
\usepackage{amsmath}
\usepackage{amssymb}
\usepackage{hyperref}
\usepackage{relsize}
\usepackage{graphicx}
\usepackage{subfig}
\usepackage[english]{babel}
\usepackage{xcolor}
\usepackage[ruled,vlined, linesnumbered]{algorithm2e}

\usepackage{comment} 
\newcommand{\ignore}[1]{}
\usepackage{color} 
\usepackage{multirow}
\usepackage[T1]{fontenc}

\makeatletter
\newcommand{\removelatexerror}{\let\@latex@error\@gobble}
\makeatother

\IEEEoverridecommandlockouts
\begin{document}

\title{PowerNet:  Multi-agent Deep Reinforcement Learning for Scalable Powergrid Control}

\author{Dong Chen$^{1}$, Kaian Chen$^{1}$, Zhaojian Li$^*{^1}$, Tianshu Chu$^2$, Rui Yao$^3$, Feng Qiu$^3$, Kaixiang Lin$^4$

\thanks{$^1$ Dong Chen, Kaian Chen and Zhaojian Li are with the Department of Mechanical Engineering, Michigan State University, Lansing, MI, 48824, USA. Email: {\tt \{chendon9,chenkaia,lizhaoj1\}@msu.edu.}}
\thanks{$^2$ Tianshu Chu is with the Department of Civil and Environmental Engineering, Stanford University, Stanford, CA 94305, USA. Email: \tt cts198859@hotmail.com.}
\thanks{$^3$ Rui Yao and Feng Qiu are with the Argonne National Laboratory, Lemont, IL 60439, USA. Email: \tt yaorui.thu@gmail.com; fqiu@anl.gov.}
\thanks{$^4$ Kaixiang Lin is with the Department of Computer Science, Michigan State University, Lansing, MI, 48824, USA. Email: \tt linkaixi@msu.edu.}
\thanks{$*$ Zhaojian Li is the corresponding author.}}

\maketitle

\begin{abstract}
This paper develops an efficient multi-agent deep reinforcement learning algorithm for cooperative controls in powergrids.
Specifically, we consider the decentralized inverter-based secondary voltage control problem in distributed generators (DGs), which is first formulated as  a cooperative multi-agent reinforcement learning (MARL) problem. We then propose a novel on-policy MARL algorithm, PowerNet, in which each agent (DG) learns a control policy based on (sub-)global reward but local states and encoded communication messages from its neighbors.
Motivated by the fact that a local control from one agent has limited impact on agents distant from it, we exploit a novel spatial discount factor to reduce the effect from remote agents, to expedite the training process and improve scalability.  Furthermore, a differentiable, learning-based communication protocol is employed to foster the collaborations among neighboring agents.
In addition, to mitigate the effects of system uncertainty and random noise introduced during on-policy learning, we utilize an action smoothing factor to stabilize the policy execution.   To facilitate training and evaluation, we develop PGSim, an efficient, high-fidelity powergrid simulation platform. 
Experimental results in two microgrid setups show that the developed PowerNet outperforms the conventional model-based control method, as well as several state-of-the-art MARL algorithms. The decentralized learning scheme and high sample efficiency also make it viable to large-scale power grids. 


\end{abstract}

\IEEEpeerreviewmaketitle

\vspace{-5pt}
\section{Introduction}
Renewable energy such as solar and wind has attracted significant research interest in  recent decades due to its great potential to reduce greenhouse gas emissions and subsequently slow global warming \cite{SOOD20191}. In a modern power grid, those renewable energy sources are present as distributed generators (DGs), along with other traditional electricity sources such as 
fossil-fuel and nuclear stations. In particular, microgrids are localized subgrids that can operate either connected to or isolated from the main grid. As microgrids can still operate when the main grid is down, they can strengthen grid resilience and service reliability. 
 There are two levels of controls in a microgrid when disconnected from the main grid: primary control and secondary control \cite{bidram2014multiobjective}. The primary control refers to the lower level control within a DG to maintain a desired voltage reference whereas the secondary control  is concerned with the cooperative generations of local references across the DGs to achieve grid-wise control objectives \cite{lai2016droop, guerrero2010hierarchical, mehrizi2009secondary, savaghebi2012secondary, shafiee2013distributed}. As the secondary control has the great potential to improve the overall grid efficiency, it will be the focus of this paper.

Existing secondary  control methods can be categorized into two main classes: centralized and distributed. A centralized controller gathers information from all DGs and then makes a collective control decision that is then sent to respective DGs \cite{vovos2007centralized,mehrizi2010potential,mnih2013playing}.While these centralized voltage control schemes show promising results, they incur heavy communication overheads and often suffer from a-single-point-of-failure \cite{wang2020data} and curse of dimensionality,  which make it impractical to deploy in today's large-scale microgrid systems.  Inspired by the idea of cooperative control in multi-agent systems, distributed control methods  have also been developed \cite{xin2010self, bidram2013distributed,shafiee2013distributed,bidram2013secondary,  bidram2014multiobjective, ding2018distributed, lu2016distributed}, which consider a more practical setup where each DG communicates with its neighboring DGs and makes a decentralized control decision based on its own state as well as the states from its neighbors, shared via local communication networks.
Conventional model-based methods generally first approximate the nonlinear heterogeneous dynamics of microgrids with a simplified linear dynamics and then develop distributed feedback controllers for the formulated tracking synchronization problem \cite{bidram2013distributed,bidram2013secondary, bidram2014multiobjective, lai2017distributed, simpson2015secondary, guo2014distributed}. 
Note that the underlying microgrid dynamics is subject to complex nonlinearity, system and disturbance uncertainty, and high dimensionality, model simplifications have to be made to enable the model-based control designs, which will inevitably negatively impact the control performance.

More recently, reinforcement learning (RL) has emerged as a promising framework to address the centralized voltage control problem due to its online adaptation capabilities and the ability to solve complex problems  \cite{lillicrap2015continuous, silver2016mastering, mnih2016asynchronous, chen2020autonomous, glavic2019deep}. 
In particular, Deep Q-network \cite{mnih2013playing} is exploited in autonomous grid operational control, which demonstrates promising performance despite the presence of load/generation variations and topological changes \cite{diao2019autonomous}. 
Duan et. al \cite{duan2019deep} uses deep deterministic policy gradient (DDPG) \cite{lillicrap2015continuous}, an off-policy RL algorithm,  to regulate the voltages of buses to the normal range by taking deterministic actions, which also shows promising performance. A two-time scale voltage controller is proposed in \cite{yang2019two}, where shunt capacitors are configured to
minimize the voltage deviations using a deep reinforcement learning algorithm.

Meanwhile, multi-agent reinforcement learning (MARL) has been greatly advanced and successfully applied to  a variety of complex systems in the past few years, including role play games like StarCraft and DOTA \cite{vinyals2019grandmaster, berner2019dota},  traffic lights control \cite{chu2019multi}, and autonomous driving \cite{shalev2016safe}.
The applications of MARL to voltage control of microgrids also exist \cite{liu2020online, cao2020multi, zhang2020deep, gao2021consensus}, which focus on autonomous voltage control (AVC), with the objective of regulating the voltage of all buses across the power systems.
In this paper, we treat the secondary voltage control in isolated microgrids with the goal of stabilizing the output voltages of the DGs at a pre-defined reference value \cite{bidram2013distributed}. We formulate the secondary voltage control (SVC) of inverter-based microgrid systems as a partially observable Markov decision process (POMDP) and develop an on-policy MARL algorithm, PowerNet, to efficiently coordinate the  DGs for cooperative controls.  We develop a decentralized, on-policy MARL approach that is both stable in training and efficient in policy learning. Comprehensive experiments are conducted and results show that our proposed PowerNet outperforms several state-of-the-art MARL algorithms, as well as a model-based approach in terms of performance, learning efficiency, and scalability. 

The main contributions and the technical advancements of this paper are summarized as follows.
\begin{enumerate}
\item The secondary voltage control of inverter-based microgrids is modeled as a decentralized, cooperative MARL problem, and a corresponding power grid simulation platform, PGSIM, is developed and open-sourced\footnote{See \url{https://github.com/Derekabc/PGSIM}}.
\item We develop an efficient on-policy decentralized MARL algorithm, PowerNet, to effectively learn a control policy by introducing a novel spatial discount factor, a learning-based communication protocol, and an action smoothing mechanism.
\item Comprehensive experiments are performed, which show that the proposed PowerNet outperforms the conventional model-based control method, as well as 6 state-of-the-art MARL algorithms, in terms of sample efficiency and voltage regulation performance.
\end{enumerate}

The remainder of the paper is organized as follows. 
The background of reinforcement learning and state-of-the-art MARL algorithms are presented in Section~\ref{sec:2}.
In Section~\ref{sec:3}, we formulate the secondary voltage control as an MARL problem and  our developed MARL algorithm, PoweNet, is detailed in Section~\ref{sec:4}. 
Experiments, results, and discussions are presented in Section~\ref{sec:5} whereas concluding remarks and future works are drawn in Section~\ref{sec:6}.

\begin{figure*}[!t]
    \centering
    \subfloat[\centering Diagram of decentralized control of microgrids.  ]{{\includegraphics[width=.42\linewidth]{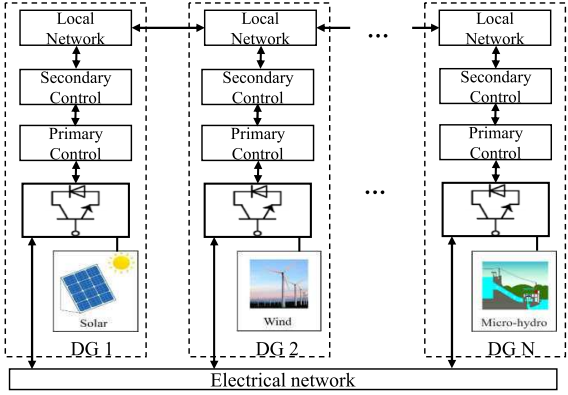} }}%
    \qquad
    \subfloat[\centering Diagram of inverter-based DG. ]{{\includegraphics[width=.48\linewidth]{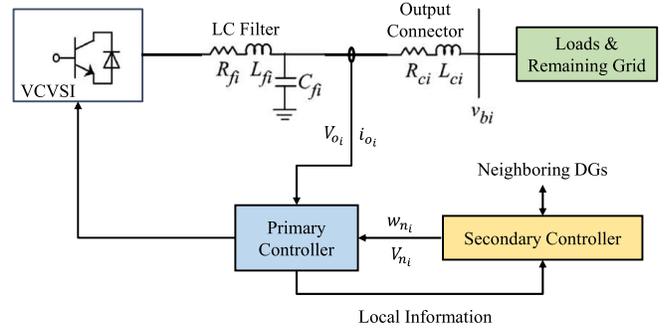}}}
    \caption{Schematic diagram of the decentralized control of microgrids and the inverter-based DG.}
    \label{fig:droop}
    \vspace{-15pt}
\end{figure*}

\section{BACKGROUND}\label{sec:2}
In this section, we review the preliminaries of reinforcement learning (RL) and several state-of-the-art MARL algorithms that will be used as benchmarks for comparison in Section~V.
\vspace{-20pt}
\subsection{Preliminaries of Reinforcement Learning (RL)}
The objective of a RL agent is to learn an optimal policy $\pi^*$ that maximizes the accumulated reward $R_t = \sum_{k=0}^{T} \gamma ^k r_{t+k}$ via  continuous interactions with the environment. Here $r_{t+k}$ is the (random) instantaneous reward at time step $t+k$ and $\gamma\in (0,1]$ is a discount factor that places more emphasis on near-term rewards. More specifically, at each discrete time step $t$, the agent takes an observation of the environment and itself $s_t\in\mathbb{R}^{n}$, executes an (exploratory) action $a_t\in\mathbb{R}^{m}$, and receives a scalar reward $r_t\in\mathbb{R}$. In general, the environment is impacted by some unknown disturbance $w_t$.  A policy, $\pi:\,\mathcal{S}\rightarrow\Pr(\mathcal{A})$, is a mapping from states to a probability distribution over the action space and it describes the agent's behavior.
If the agent can only observe $o_t$, a partial description of the  state $s_t$, the underlying dynamics becomes a partially observable Markov Decision Process (POMDP) and the goal is then to learn a policy that maps from the observation $o_t$ to an appropriate action. 

 The value function of a state $s$ under policy $\pi$, $V^{\pi}(s)$, is defined as the expected return starting from $s$ and following policy $\pi$, i.e., $V^{\pi}(s_t) = E_{\pi}{[R_t | {s_t = s}]}$. The state-action value function (or Q-function), $Q^{\pi}(s,a)$, is defined as the expected return starting from state $s$, taking  first action $a$, and then following policy $\pi$ afterwards. The optimal Q-function for a state-action pair ($s,a$) is given by $Q^{*}(s,a) = \max_{\pi} Q^{\pi}(s,a)$. Note that the knowledge of the optimal Q-function induces an optimal policy by $\pi^*(s)=\arg\max_{a}Q^{*}(s,a)$. Many times the agent's policy is parameterized by some parameters $\theta$ and  the goal is to learn the values of parameters $\theta$ that maximize the reward function. In actor-critic (A2C) algorithms \cite{mnih2016asynchronous}, two networks are employed:  a critic network parameterized by $\omega$ to learn the value function $V_{\omega}^{\pi_\theta}(s_t)$ and an actor network $\pi_{\theta}(a_t|s_t)$ parameterized by $\theta$ to update the policy distribution in the direction suggested by the critic network by:
\begin{equation}\label{eqn:advantagepolicygradient}
\theta \leftarrow \theta + E_{\pi_{\theta}} \left[ \Big (\nabla_{\theta} \log \pi_{\theta}(a_t|s_t) \Big)
 A_t \right].
\end{equation}
where $A_t= Q^{\pi_\theta}(s,a) - V_{\omega}^{\pi_\theta}(s_t)$ \cite{mnih2016asynchronous} is the advantage function used to reduce the sample variance. The value function parameters are then updated by minimizing the following square error loss: 

\begin{equation}\label{eqn:valueloss}
\min_{\omega} E_{\mathcal{D}}\Big (R_t + \gamma V_{\omega'} (s_{t+1}) - V_{\omega}(s_t)\Big )^2.
\end{equation} 
where $\mathcal{D}$ denotes the experience replay buffer that collects previously encountered trajectories and $\omega'$ denotes the parameters obtained from earlier iterations as a target network \cite{mnih2013playing}.

\vspace{-10pt}
\subsection{MARL and State-of-the-Art}
 MARL is an extension of the single-agent agent and is concerned with systems with multiple agents/players.  Great progress has been made in the past decade with many successful applications \cite{vinyals2019grandmaster, berner2019dota, chu2019multi, shalev2016safe}. Independent Q-learning (IQL) \cite{tan1993multi} is a natural extension from single-agent RL, allowing each agent to learn independently and simultaneously while viewing other agents as part of the environment. Although fully scalable, IQL suffers from non-stationarity and partial observability. To address the non-stationary issue of IQL,  FPrint \cite{foerster2017stabilising} exploits the technique of importance sampling and includes low-dimensional fingerprints to distinguish obsolete data in the experience buffer. 
MADDPG \cite{DBLP:journals/corr/LoweWTHAM17} extends single-agent actor-critic methods to multiple agents where the critic network is augmented with additional information of other agents' policies during training.
Furthermore, ConseNet \cite{zhang2018fully} proposes an actor-critic-based MARL algorithm with fully decentralized critic networks through a consensus update in the communication network. 
An  emergent  paradigm  in  MARL  is \textit{learn  to  communicate}, in  which  agents  automatically  learn  a  communication  protocol  to  maximize  their  shared  utility. For example, in DIAL \cite{foerster2016learning}, agents receive messages from their  neighbors through a differentiable communication protocol, in which the message is encoded and summed with other input signals at the receiver side.
Another framework is CommNet \cite{sukhbaatar2016learning}, which learns the communication protocol by calculating the mean of all messages instead of encoding them. Note that for both DIAL and CommNet, information losses  inevitably incur as they involve summation of input signals \cite{chu2020multi}.

Inspired by the aforementioned works, in this paper, we develop a new on-policy MARL algorithm, PowerNet, for secondary voltage control. It exploits the \textit{centralized learning decentralized execution} structure as in \cite{lowe2017multi}. To facilitate cooperation among the neighboring agents, instead of summing the messages, we concatenate and encode the communication messages. Novel techniques of spatial discount factor and action smoothing are also incorporated to improve learning efficiency and control performance. More details are presented in Section IV.

\section{Problem formulation}\label{sec:3}
Voltage-controlled voltage source inverter (VCVSI) is widely used in DGs to provide fast voltage/frequency support \cite{bidram2013distributed}. 
A typical VCVSI with decentralized control architecture is illustrated in Fig.~\ref{fig:droop}a, where a \textit{secondary controller} is employed for each DG to coordinate with neighboring DGs and dynamically generate voltage references. The obtained reference is then used by the lower-level \textit{primary controller} as the reference to track. The overall control objective is to maintain the voltage and frequency of all DGs to the reference value despite the disruptions in the power network and tracking imperfections in primary control \cite{bidram2013distributed, bidram2013secondary, ning2020distributed}. More specifically, as shown in Fig. \ref{fig:droop}b,  the primary controller of each DG $i$, $i=1,\cdots,N$, consumes the frequency and voltage references, $w_{n_i}$ and $V_{n_i}$, from the secondary controller, and regulate the output voltage $V_{oi}$ and frequency to a desired reference.  This is typically achieved via the active and reactive-power droop techniques \cite{bidram2013distributed, guerrero2010hierarchical} without communications among DGs. The readers are referred to \cite{bidram2013distributed, bidram2013secondary} for the underlying system dynamics, which is exploited to develop our power grid simulation platform, PGSIM\footnote{See \url{https://github.com/Derekabc/PGSIM}}. The objective of the SVC is to coordinate with other DGs and generate reference signals $V_{n_i}$ to synchronize the voltage magnitude of DG $i$ to the reference value, in the presence of power disturbances and primary control imperfections.

While model-based secondary control approaches exist \cite{bidram2014multiobjective,bidram2013distributed, bidram2013secondary}, the performance is generally unsatisfactory due to system simplifications made to address nonlinearity and  uncertain disturbances. In this paper, we develop a model-free, MARL-based approach. Specifically, we model a microgrid as a multi-agent  network: $\mathcal{G} = (\text{\larger[2]$\nu$}, \text{\larger[2]$\varepsilon$})$, where each agent $i  \in \text{\larger[2]$\nu$}$  communicates with its neighbors $\mathcal{N}_i:=\{j|\varepsilon_{ij}\in\large\text{\larger[2]$\varepsilon$})\}$. 
Let $S := \times_{i \in \text{\larger[2]$\nu$}} S_i$ and $\mathcal{A} := \times_{i \in \text{\larger[2]$\nu$}} \mathcal{A}_i$
be the global state space and action space that represent, respectively, the aggregated state information and concatenated controls of all DGs, the underlying dynamics of the microgrid can be characterized by the state transition distribution $\mathcal{P}$: $\mathcal{S} \times \mathcal{A} \times \mathcal{S} \rightarrow [0, 1]$. We consider a decentralized MARL framework to achieve scalable power grid control. More specifically, each DG only communicates with its neighbors and makes a control decision based on these observations. As each agent $i$ (DG $i$) only observes part of the environment (states of itself and its neighbors),  this leads to a POMDP \cite{hausknecht2015deep}.  At each time step t, each agent $i$ makes an observation $o_{i,t}\in \mathcal{O}_{i}$, takes an action $a_{i,t}$, and then receives the next observation ${o_{i,t+1}}$ and a reward signal $r_{i, t} := \mathcal{O}_t \times \mathcal{A}_t \rightarrow \mathbb{R}$. The goal is to learn an optimal decentralized policy $\pi_i := \mathcal{O}_{i} \times \mathcal{A}_{i} \rightarrow [0, 1]$ such that the expected future rewards are maximized. We solve the above problem through MARL and we define the key elements in the considered  POMDP in the following.

\begin{itemize}
\item \textit{Action space}: the control action for each DG is  the secondary voltage control set point $V_{n_i}$. In this paper, we use 10 discrete actions that are evenly spaced between $1.00~pu$ and $1.14~pu$. The overall action of the microgrid is the joint actions from all DGs, i.e., $a=v_{n_1}\times v_{n_2}\times\cdots \times v_{n_N}$.

\item \textit{State space}: the state of each DG $i$ is chosen as  $s_{i,t} = (\delta_i, P_i, Q_i, i_{odi}, i_{oqi}, i_{bdi}, i_{bqi},v_{bdi}, v_{bqi})$ to characterize the operations of the DG, where $\delta_i$ is the measured reference angle frame; $P_i$ and $Q_i$ denote the active and reactive power, respectively; $i_{oqi}$, $i_{oqi}$, $i_{bdi}$ and $i_{bqi}$ represent the output d-q currents of the DG $i$ and the directly connected buses, respectively; and $v_{bdi}$ and $v_{bqi}$ are the output d-q voltages of the connected bus, respectively. The entire state of the microgrid system is the Cartesian product of the individual states, i.e.,  $S(t)=s_{1,t}\times\cdots\times s_{N,t}$.  

\item \textit{Observation space}: we assume that each DG can only observe its local state as well as messages from its neighbors, i.e.,  {}$o_{i,t} = s_{i,t} \cup m_{i,t}$, where $m_{i,t}$ is the communication message received from its neighboring agents  $j \in \mathcal{N}_i$ and will be detailed in Section IV.

\item \textit{Transition probabilities}: the transition probability $\mathcal{T}(s'|s,a)$ characterizes the dynamics of the microgrid. We follow the models in  \cite{bidram2013distributed, bidram2013secondary} to build our simulation platform but we do not exploit any prior knowledge of the transition probability as our MARL is model free.  

\item \textit{Reward function}: we design the following reward function to promote the DGs to quickly converge to reference voltages (e.g., $1~pu$):  
\begin{equation}\label{eqn:reward_fn}
r_{i,t} \triangleq 
    \begin{cases}
          0.05 - |1-v_i|, & \text{$v_i \in [0.95,1.05]$,} \\
          - |1-v_i|, & \text{$v_i \in [0.8, 0.95] \; \cup \; [1.05, 1.25]$,} \\
          -10, & \text{Otherwise}.
    \end{cases}
\end{equation}
where $r_{i,t}$ is the reward of agent $i$ at time step $t$. More specifically, we divide the voltage range into 3 operation zones similar to \cite{wang2020data}: normal zone ($[0.95, 1.05]~pu$), violation zone ($[0.8, 0.95]\; \cup\;[1.05, 1.25]~ pu$), and diverged zone ($[0, 0.8]\; \cup\; [1.25, \infty]~pu$). With the formulated reward, DGs with diverged voltages or have no powerflow solution will receive large penalty, while DGs with voltage close to $1~pu$ obtain positive rewards.
\end{itemize}

\textit{\bf Remark 1.} The formulation of regulating the voltages of DGs follows from literature such as \cite{bidram2013distributed, bidram2014multiobjective, mustafa2019detection}. This formulation can regulate the DG voltages but the bus voltages may still go beyond the normal range. An alternative formulation for regulating bus voltages can be addressed similarly by considering a different reward function, the demonstration of which can be found at the site\footnote{See \url{https://github.com/Derekabc/PGSIM/tree/R2}}.

\section{PowerNet for secondary voltage control}\label{sec:4}
In this section, we present a novel decentralized MARL algorithm, PowerNet, to efficiently solve the POMDP formulated above. The proposed PowerNet extends the independent actor-critic (IA2C) to deal with multiple cooperative agents by fostering collaborations between neighboring agents, which is enabled by the following three novel characteristics: 1) a learning-based differentiable communication protocol, to efficiently promote cooperation between neighboring agents; 2) a novel spatial discount factor, to mitigate partial observability and enhance learning stability; and 3) an action smoothing scheme, to mitigate the effects of system uncertainty and random noise introduced during on-policy learning. We next explain those features in more detail.

\subsubsection{Differentiable communication protocol}

\begin{figure}
\centering
\includegraphics[width=0.48\textwidth]{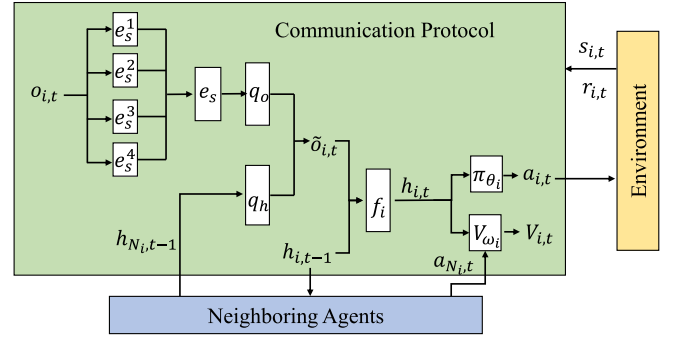}
\caption{Overview of the proposed communication protocol.}
\label{fig: system_diag}
\vspace{-5pt}
\end{figure}

We consider a decentralized MARL framework where each agent (DG) can communicate with its neighbors and exchange necessary information such as the encoded states. In contrast to non-communicative MARL algorithms (e.g., IA2C, FPrint and ConseNet) that generally suffer from slow convergence and unsatisfactory performance \cite{chu2020multi}, we incorporate information from neighboring agents to enhance the learning efficiency. As illustrated in Fig. \ref{fig: system_diag}, at each time step $t$, agent $i$ updates its hidden state $h_{i,t}$ as:
\begin{equation}\label{eqn:hidden_info}
h_{i, t} = f_i (h_{i, t-1}, q_o (e_s(o_{i,t})), q_h (h_{\mathcal{N}_{i,t-1}})).
\end{equation}
where $h_{i, t-1}$ is the encoded hidden state from last time step; $o_{i,t}$ is agent $i$'s observation made at time $t$, i.e., its internal state \ignore{and states from its neighbors};  $h_{\mathcal{N}_{i,t-1}}$ is the concatenated hidden state from its neighbors;  $e_s$, $q_o$, and $q_h$ are differentiable message encoding and extracting functions, where one-layer fully connected layers with 64 neurons are used; and $f_i$ is the encoding function for the hidden states and communication information, where we use a Long Short Term Memory (LSTM \cite{hochreiter1997long}) network with a 64-neuron hidden layer to improve the observability \cite{hausknecht2015deep} and better utilize the past hidden state $h_{i, t-1}$ information.

\ignore{Instead of using low-dimensional indicators as in \cite{foerster2017stabilising}, we include the neighbor's complete states into the local observation $o_{i, t} = s_{i, t} \cup s_{\mathcal{N}_i, t}$ to improve the agent's observability and use a network to learn an appropriate representation automatically.} 

To improve the scalability and robustness, we  regroup the observation $o_{i, t}$ according to their physical units. For instance, the observation $o_{i,t}$ is divided into four groups: ${o}_{i,t}^1  \cup {o}_{i,t}^2 \cup {o}_{i,t}^3 \cup {o}_{i,t}^4$ according to their units, i.e., reference angle frame, power, voltage, and current. These regrouped sub-observations are encoded independently and then concatenated as $e_s (o_{i,t})  = cat(e_{s}^1({o}_{i,t}^1), e_{s}^2({o}_{i,t}^2), e_{s}^3({o}_{i,t}^3), e_{s}^4({o}_{i,t}^4))$, where $e_{s}^j$, j=1,2,3,4, are one-layer fully connected encoding layers. 

The received communication message $m_{i,t}$ of the $i$th agent is \ignore{a combination of internal states and} the encoded hidden states of its neighbors, i.e., $m_{i,t} =  h_{\mathcal{N}_i, t-1}$ with $h_{\mathcal{N}_i, t-1}$ being the hidden states of agent $i$'s neighbors at time $t-1$. Note that since the hidden state $h_{t-1}$ is neural encoded, these messages are much more secure as compared to directly sending the raw states. 
The encoded observation $e_s(o_{i,t})$ and the neighbors' hidden states $h_{\mathcal{N}_{i,t-1}}$ are extracted by $q_o$ and $q_h$, respectively. 
We then concatenate the encoded message as $\tilde{o}_{i,t} = cat(q_o (e_s(o_{i,t})), q_h(h_{\mathcal{N}_{i,t-1}}))$. 
Concatenation operation is shown in \cite{chu2020multi} to have better performance as compared to the summation operation used in DIAL and CommNet on reducing information loss. The hidden state is then obtained by using the LSTM $f_i$ to encode $\tilde{o}_{i,t}$ and $h_{i, t-1}$.

The hidden state $h_{i, t}$ obtained from (\ref{eqn:hidden_info}) is then used in the actor and critic networks to generate random actions and predict the value functions, respectively, i.e.,  $\pi_{\theta_i}(\cdot | h_{i, t})$ and $V_{\omega_i} (h_{i,t})$.  Inspired by MADDPG \cite{DBLP:journals/corr/LoweWTHAM17}, we also include the neighbors' action information into the critic network  $V_{w_i}(h_{i,t}, a_{\mathcal{N}_i ,t})$ to enhance the training.
In this paper, we use a discrete action space and the action is sampled from the last Softmax layer as $a_{i,t} \sim \pi_{\theta_i}(\cdot | h_{i,t})$. 
We adopt the centralized training and decentralized execution scheme \cite{chu2020multi,chu2019multi}, where each agent has its own actor and critic networks, and their policies are updated independently instead of updating in a consensus way \cite{zhang2018fully} that may hurt the convergence speed. 

\subsubsection{Spatial discount factor}
As mentioned above, the goal of cooperative MARL is to maximize the shared global reward $R_{g,t} = \sum_{i \in \text{\larger[2]$\nu$}} R_{i,t}$, where $R_{i,t} = \sum_{k=0}^{T} \gamma ^k r_{i,t+k}$ denotes the cumulative reward for agent $i$. For each agent, a natural choice of the reward is the instantaneous global reward, i.e., $\sum_{i=1}^{N} r_{i,t}$. However, this scheme can lead to several issues. First, aggregating global reward can cause large latency and increase the communication overheads. Second, the single global reward leads to the credit assignment problem~\cite{sutton2018reinforcement}, which would significantly impede the learning efficiency and limit the number of agents to a small size.
Third, as an agent is typically only slightly impacted by agents distant from it, using the global reward to train the policy for each agent can lead to slow convergence. To address the aforementioned issues of using a global reward, in this paper, we employ a novel  spatial discount factor to address the above challenges. Specifically, each agent $i, \,i=1,\cdots,N$, utilizes the following reward:
\begin{equation}\label{eqn:distance_return}
R_{i, t} = \sum_{k=0}^{T} \gamma ^k \sum_{j \in \nu} \alpha({d_{i,j}}) r_{j, t+k}.
\vspace{-5pt}
\end{equation}
where $\alpha(d_{i,j}) \in [0, 1]$ is the spatial discount function with $d_{i,j}$ being the distance between agent $i$ and $j$. The distance $d_{i,j}$ can be an Euclidean distance characterizing the physical distance between the two agents or the distances between two vertices in a graph (i.e., number of shortest connecting edges).  Note that the new reward defined in (\ref{eqn:distance_return}) characterizes a whole spectrum of reward correlation, from local greedy control (When $\alpha(d_{i,j\neq i})=0$ and $\alpha(d_{i,i})=1$), to a global reward (when $\alpha\equiv1$). Note that there are different choices of the spatial discount function. For example, one can choose 
\begin{equation}
\alpha(d_{i,j})=   \begin{cases}
          1, & \mbox{if $d_{ij}\leq D$,} \\
         0, & \mbox{Otherwise}.
    \end{cases}
\vspace{-5pt}
\end{equation}
where $D$ is the distance threshold that defines an ``effective distance'' near the considered agent. The threshold $D$ can incorporate factors such as communication speed and overhead. In this paper, we adopt $\alpha(d_{i,j})=\alpha^{d_{i,j}}$ with $\alpha\in(0,1]$ being a constant scalar and a hyperparameter to be tuned.

As a result, the gradient computation in Eqn. (\ref{eqn:advantagepolicygradient}) becomes:
\begin{equation}\label{eqn:advantagepolicygradient_MARL}
\nabla_{\theta_i} {J(\theta_i)} = E_{\pi_{\theta}} \left[ \Big (\nabla_{\theta_i} \log \pi_{\theta_i}(a_{i, t}|o_{\nu_i, t}) \Big)
 A_{i, t} \right].
\vspace{-5pt}
\end{equation}
where $A^{\pi_{\theta_i}}_{i, t} = Q^{\pi_{\theta_i}}(s,a) - V^{\pi_{\theta_i}}(s)$ is the advantage function, $Q^{\pi_{\theta_i}}(s,a) = E_{\pi_{\theta_i}} (R_{i,t} | s_t =s, a_{\nu_i, t} = a_{\nu_i})$ is the state-action value function,  and $V^{\pi_{\theta_i}}(s) =V^{\pi_i}(s, a_{\mathcal{N}_i})$ is the state value function.
We optimize the parameter of critic network $V_{w_i}$ as follows:
\begin{equation}\label{eqn:valueloss_MARL}
\min_{\omega_i} E_{\mathcal{D}} \Big [\sum_{j \in \nu} \alpha^{d_{ij}} r_{j, t} + V_{\omega'_i}(o_{\nu_i, t}) - V_{\omega_i}(o_{\nu_i, t})\Big ]^2.
\vspace{-5pt}
\end{equation}
Minibatches of sampled trajectory are used to update the network parameters using Eqn.~\ref{eqn:advantagepolicygradient_MARL} and Eqn.~\ref{eqn:valueloss_MARL} to reduce the variance \cite{chu2019multi}.

\subsubsection{action smoothing}
Note that our proposed PowerNet is an on-policy MARL algorithm that needs to sample  stochastic actions at each time step, which sometimes can lead to noisy action samples with large fluctuations, even after the algorithm converges. These action fluctuations can cause undesirable system perturbations. Towards this end, we introduce an action smoothing scheme  to smoothen the sampled action  $a_{i,t} \sim \pi_{\theta_i}(\cdot | h_{i,t})$ before execution as:
\begin{equation}\label{eqn:action_smoothing}
    a_{i,t} \leftarrow
    \begin{cases}
          a_{i,t}, & t=1; \\
         \rho a_{i, t-1}   + (1- \rho) a_{i,t},  & t>1. 
    \end{cases}
\vspace{-5pt}
\end{equation}
where $\rho \in [0, 1]$  is the discount factor; $a_{i, t-1}$ is the action taken at time $t-1$. This scheme is also known as exponential moving average \cite{gardner1985exponential}. When $\rho$ is chosen as 0, there is no smoothing on actions and the agent will directly take actions sampled from the policy $\pi_{\theta_i}$. When $\rho=1$, the new action $a_{i,t}$ will not be updated  and the agent keeps using the previous action.  We specify an action smoothing window $T_w$ in which the past actions are buffered and used for smoothing. This time window size needs to be properly chosen as too large $T_w$ will consider too many obsolete actions that may cause the agent to  react slowly to abrupt changes whereas too small $T_w$ may have limited smoothing effect. In this paper, $\rho$ and $T_w$ are treated as hyperparameters and are tuned using cross-validations.

The complete PowerNet  algorithm is shown in Algorithm~\ref{powernet}. The hyperparameters include: the distance discount factor $\alpha$, the action smoothing factor $\rho$, the (time)-discount factor $\gamma$, the learning rates of the actor network $\eta_w$ and critic network $\eta_{\theta}$, the total number of training steps $M$, the epoch length $T$, the action sample window $T_w$, and the batch size $N$. More specifically, the agents interact with the environment for thousands of epochs  (Lines 2--29). In Lines 4--6, each agent collects and sends the communication message $m_{i,t}$ to its neighbors. After that, each agent combines, regroups and encodes its observations (Line 8). 
The agents then update their hidden states and actor networks used to sample actions (Lines 9--10). In Lines 12--15, the state value is updated and actions are smoothed before execution. After interacting with the environment (Lines 16), agents transition to the next state and receive an immediate reward (Lines 17), which will be added to an on-policy experience buffer (Lines 18). In Lines 20--26, the parameters of the actor and critic networks are updated using the collected trajectories sampled from the on-policy experience buffer after the end of each episode. The DONE signal is flagged if either the episode is completed or there is no powerflow solution. After receiving the DONE flag, all agents are reset to their initial states  to start a new epoch  (Lines 28).

\begin{figure}
\removelatexerror
\scalebox{0.85}{
\begin{algorithm*}[H]
\SetAlFnt{\small}
    \SetKwInOut{Parameter}{Parameter}
    \SetKwInOut{Output}{Output}
\caption{PowerNet for Secondary Voltage Control}
\label{powernet}
\SetAlgoLined
\Parameter{$\alpha, \rho, \gamma, \eta_w, \eta_{\theta}, T, M, T_w$.}
\Output{$\theta_i, w_i, i \in \text{\larger[2] $\nu$}$.}
\hrule
\vspace{0.2em}
{\bf initialize} $s_0, h_{-1},t \leftarrow 0, \mathcal{D} \leftarrow \emptyset$;\\
\For{$j = 0$ to $M-1$}{
    	\For{$t=0$ \text{to} $T-1$}{
        	\For{$i \in \text{\larger[2]$\nu$}$}{
        	     send $m_{i,t} = h_{i,t-1}$}
        	    
        	  \For{$i \in \text{\larger[2]$\nu$}$}{
	            get $\tilde{o}_{i,t} = cat(q_o (e_s(o_{i,t})), q_h(h_{\mathcal{N}_{i,t-1}}))$\\
	            get $h_{i,t} = f_i (h_{i, t-1}, \widetilde{o}_{i,t}), \pi_{i,t} \leftarrow \pi_{\theta_i}(\cdot | h_{i,t})$ \\
	            update $a_{i,t}\sim \pi_{i,t}$}
	            
	          \For{$i \in \text{\larger[2]$\nu$}$}{
            	     update $v_{i,t} \leftarrow V_{w_i}(h_{i,t}, a_{\mathcal{N}_i ,t})$\\
            	     update $a_{i,t}$ according to Eqn. (\ref{eqn:action_smoothing})\\
            	     execute $a_{i,t}$}
            	    
              {simulate $\{s_{i, t+1}, r_{i,t}\}_{i \in \text{\larger[2]$\nu$}}$}\\
        	  {update $\mathcal{D} \leftarrow \{ (o_{i,t}, a_{i,t}, r_{i,t}, v_{i,t})\}_{i \in \text{\larger[2]$\nu$}}$}\\
        	  {update $t \leftarrow t+ 1, j \leftarrow j+ 1$ }
        	  
        	  \If{DONE}{
        	   	\For{$i \in \text{\larger[2]$\nu$}$}{
            	    {update $\theta_i \leftarrow \theta_i + \eta_{\theta_i} \nabla_{\theta_i} {J(\theta_i)}$}\\
            	    {update $w_i \leftarrow w_i + \eta_{w_i} \nabla_{\omega_i} {V(\omega_i)}$}}}
    	    { initialize $\mathcal{D} \leftarrow \emptyset$}
    	}
    	{update $s_0, h_{-1},t \leftarrow 0$}}
\end{algorithm*}
}
\vspace{-10pt}
\end{figure}

\section{EXPERIMENT, RESULTS, And DISCUSSIONS}\label{sec:5}
\begin{figure*}[!ht]
  \centering
  \includegraphics[width=0.8\textwidth]{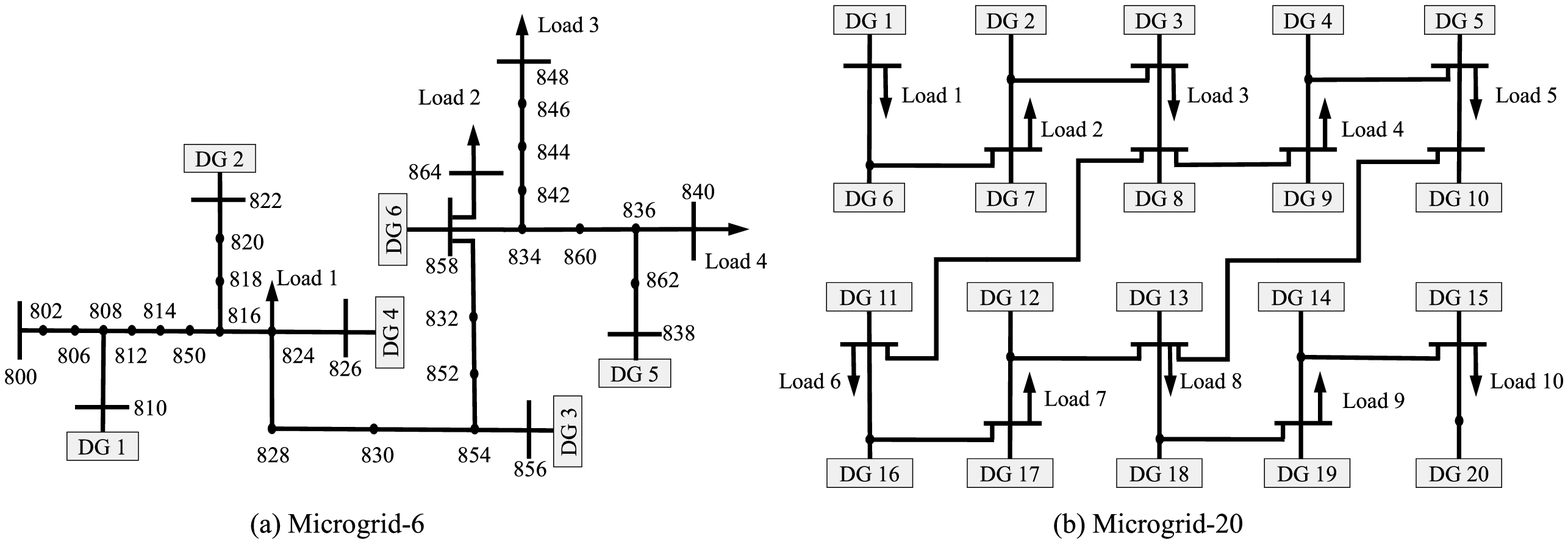}
  \caption{System diagrams of the two microgrid simulation platforms: (a) microgrid-6; and (b) microgrid-20.}
  \label{fig: microgrid}
  \vspace{-10pt}
\end{figure*}

In this section, we apply the developed PowerNet to two microgrid systems, IEEE 34-bus test feeder  with 6 distributed DGs (microgrid-6, Fig. \ref{fig: microgrid}a) and a larger-scale microgrid system with 20 DGs ( microgrid-20, Fig. \ref{fig: microgrid}b). The simulation platform we developed is based on the line and load specifications detailed in \cite{li2010consensus} and \cite{mustafa2019detection}, respectively. Furthermore, to better represent the real-world power systems, we add random load variations across the entire microgrid with $\pm 20\%$ perturbations from its nominal values specified in \cite{mustafa2019detection}. We also add random disturbances in the range of $\pm 5\%$ of the nominal values for each load at every simulation step to simulate the random disruptions in the real-world power grid. The DGs are controlled with a sampling time of 0.05s and each DG can communicate with its neighbors via local communication edges. The lower level primary control is realized through  \cite{bidram2013distributed}.
The experiments have been performed in a Ubuntu 18.04 server with AMD 9820X processor and 64 GB memory. 

Through cross-validation, the spatial discount factor $\alpha$ is chosen as 1.0 and 0.7 for microgrid-6 and microgrid-20, respectively. It is reasonable to have a larger spatial discount factor in the microgrid-6 system as agents are  strongly and tightly connected in that small microgrid system. A smaller $\alpha$ in the larger-scale microgrid-20 system is advantageous to reduce the effect of remote agents, but a too small $\alpha$ will cause the agents to ignore their effect on the neighboring  agents. For example, if we choose $\alpha=0$, the expected reward for agent $i$ at time step t will become $R_{i} = \sum_{t=1}^{T} \gamma ^t  r_{i, t}$, which cannot ensure a maximum global reward as each agent update its own policy greedily.

For the action smoothing factor $\rho$, we choose $\rho = 0.5$ for microgrid-6 and $\rho = 0.4$  for microgrid-20. As microgrid-20 is more complicated than the microgrid-6, we choose a larger $\rho$, considering that a small $\rho$ will result in a slow response to sudden voltage violations. We choose the sample window $T_w = 2$  for both microgrid systems.

\subsection{Compared to Centralized Control}
\begin{figure*}[!ht]
  \centering
  \includegraphics[width=0.75\textwidth]{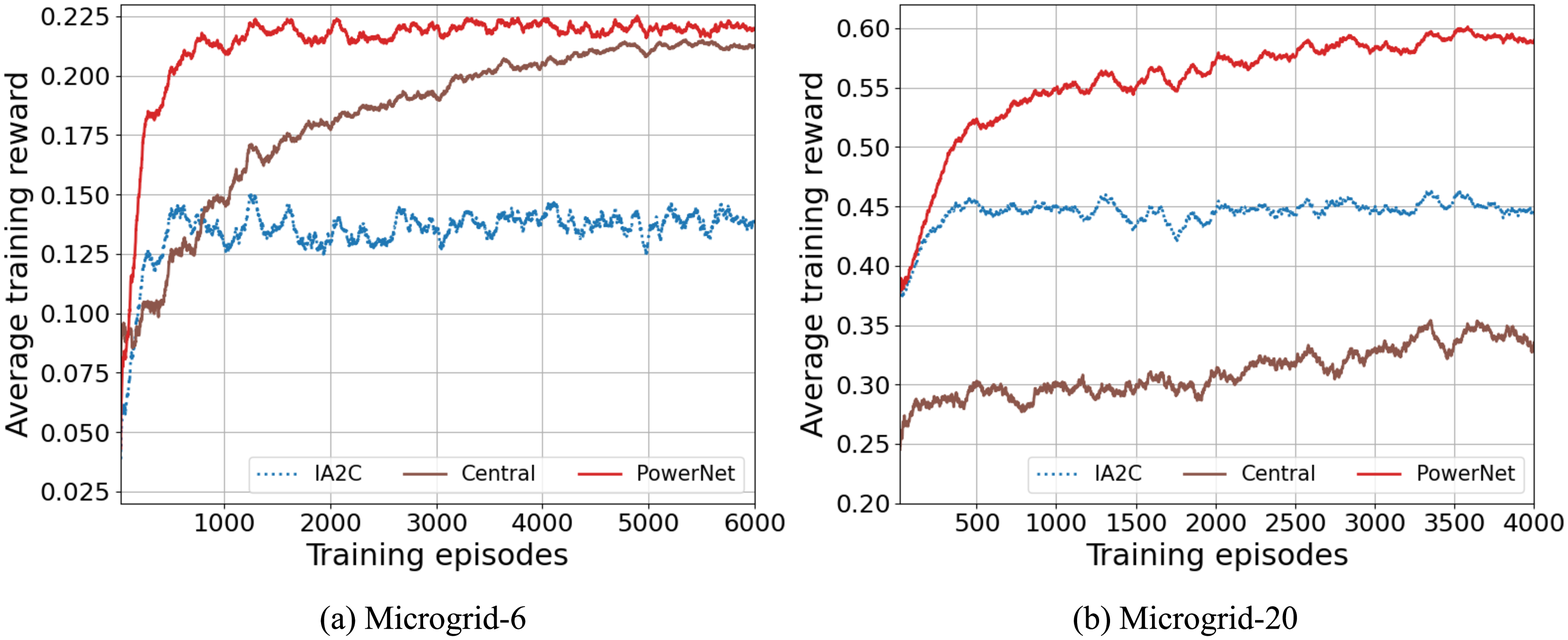}
  \caption{MARL training curves compared with PPO for (a) microgrid-6 and (b) microgrid-20 systems. The lines show the average reward per training episode which are smoothed over the past 100 episodes.}
  \label{fig:vs_central}
  \vspace{-5pt}
\end{figure*}

To compare PowerNet with a Centralized MARL approach, we apply PPO \cite{schulman2017proximal},  a state-of-the-art centralized actor-critic RL approach where the control decisions of all agents are made by a centralized RL agent that uses the global state as input, and compare its performance with PowerNet. As shown in Fig.~\ref{fig:vs_central},  PowerNet has a faster convergence speed and a better overall training performance than the centralized approach for both microgrids. Note that PPO has  poor convergence in the larger-scale microgrid with 20 DGs. This is not surprising as the centralized approach has been known to suffer from the curse of dimensionality and poor scalability. 
The input dimension for PowerNet stays unchanged as network size increases whereas centralized control method quickly becomes infeasible for large-scale networks as shown in Table \ref{tb:vs_central} .

\vspace{-8pt}

\begin{table*}[!ht]
\renewcommand{\arraystretch}{1.3}
\centering
\caption{Performance comparison between trained PowerNet and PPO.}
\label{tb:vs_central}
\begin{tabular}{c|c|c|c|c|c|c}
\hline
\multirow{2}{*}{} & \multicolumn{3}{c|}{Microgrid-6}                                                   & \multicolumn{3}{c}{Microgrid-20}                                                  \\ \cline{2-7} 
                  & Converge Time & \multicolumn{1}{l|}{Input Size} & \multicolumn{1}{l|}{Performance} & Converge Time & \multicolumn{1}{l|}{Input Size} & \multicolumn{1}{l}{Performance} \\ \hline
PPO       & 5.25 h        & 54                              & 0.21                             &  not converge         & 180                                &  0.44                             \\ \hline
PowerNet          & 0.90h & 9                            &  0.22                           & 7.28h         & 9                               & 0.60                             \\ \hline
\end{tabular}
\end{table*}

\subsection{Compared to State-of-the-art Benchmarks}
We then compare PowerNet with several state-of-the-art benchmark MARL algorithms (IA2L \cite{lowe2017multi}, FPrint \cite{foerster2017stabilising}, ConseNet \cite{zhang2018fully}, DIAL \cite{foerster2016learning}, MADDPG \cite{DBLP:journals/corr/LoweWTHAM17} ,and CommNet \cite{sukhbaatar2016learning}), as well as the conventional model-based method \cite{lou2016distributed}, to demonstrate its effectiveness.  
We train each model over 10,000 episodes, with $\gamma=0.99$, minibatch size $N = 20$, actor learning rate $\eta_{\theta} = 5 \times 10^{-4}$, and critic learning rate $\eta_w = 2.5\times 10^{-4}$. To ensure fair comparison, each episode generates different random seeds and in each episode the same random seed is shared across different algorithms to guarantee the same training/testing environment. We control the agents every $\Delta T=0.05$ seconds and one episode lasts for $T=20$ steps.

\begin{figure*}[!ht]
  \centering
  \includegraphics[width=0.76\textwidth]{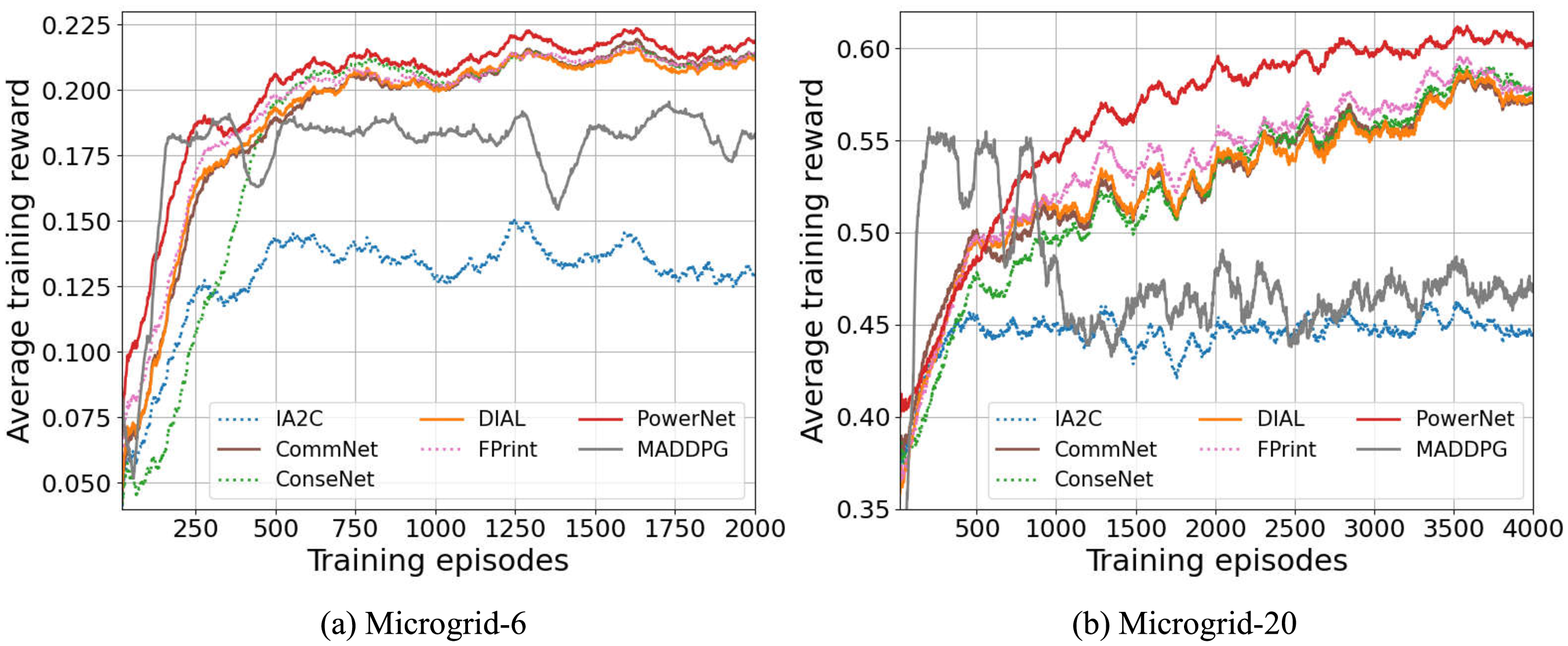}
  \caption{MARL training curves for (a) microgrid-6 and (b) microgrid-20 systems. The lines show the average
reward per training episode which are smoothed over the past 100 episodes.}
  \label{fig:overall_performance}
  \vspace{-15pt}
\end{figure*}

Fig.~\ref{fig:overall_performance} shows the training curve for all MARL algorithms in microgrid-6 and microgrid-20 systems.  To better visualize the training processes, we only show the first 2,000 and 4,000 training episodes for microgrid-6 and microgrid-20 systems, respectively. It is clear that in both microgrid-6 and microgrid-20, PowerNet outperforms all state-of-the-art MARL algorithms in terms of convergence speed. This is due to the proposed communication protocol structure and suitable spatial discount factor, which help improve sample efficiency and speed up the learning. 
In the more challenging microgrid-20 system, PowerNet shows its greater advantages of sample efficiency as seen from fastest convergence speed and best average episode reward compared to other algorithms (see Fig.~\ref{fig:overall_performance}b).

\begin{figure*}[!ht]
  \centering
  \includegraphics[width=0.9\textwidth]{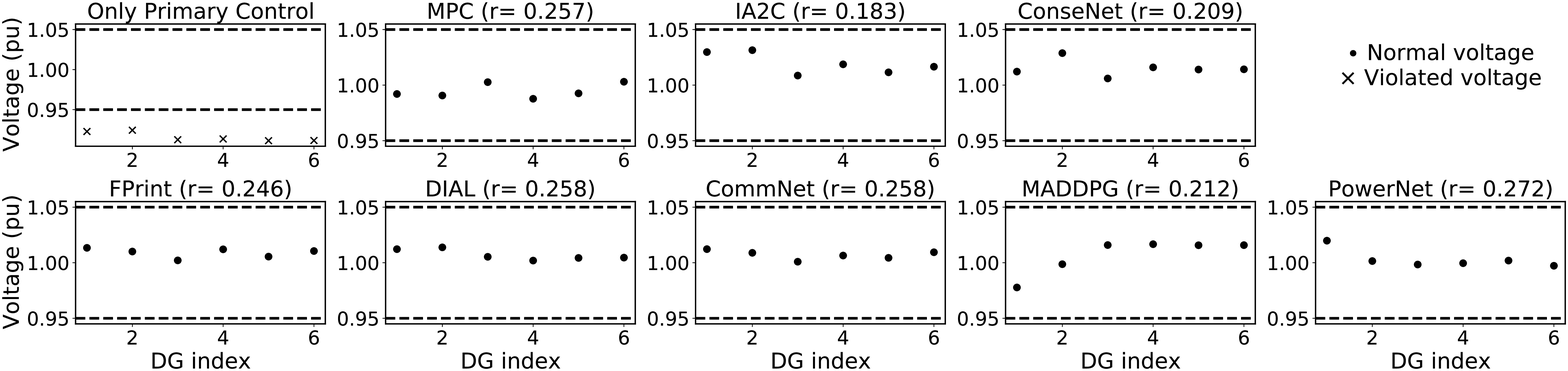}
  \caption{Execution performance on voltage control in microgrid-6. }
  \label{fig:voltage6}
  \vspace{-10pt}
\end{figure*}

\begin{figure*}[!ht]
  \centering
  \includegraphics[width=0.9\textwidth]{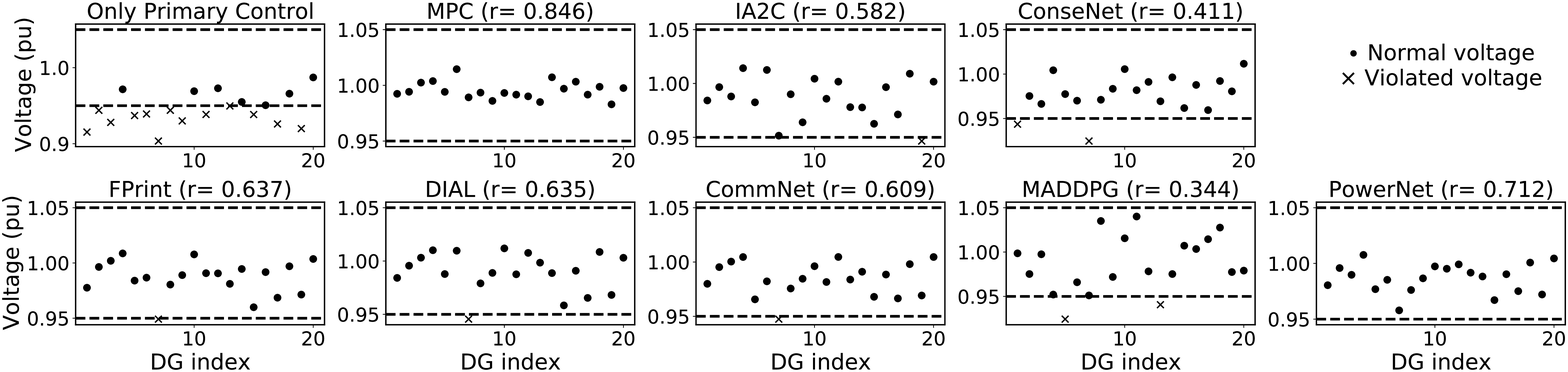}
  \caption{Execution performance on voltage control in microgrid-20.}
  \label{fig:voltage20}
  \vspace{-10pt}
\end{figure*}

To better show the voltage control performance, we compare the voltage control results between \textit{only primary controller} and \textit{with secondary controller} under heavy load variations (25\%) as shown in Fig.~\ref{fig:voltage6} (for microgrid-6) and Fig.~\ref{fig:voltage20} (for microgrid-20).
The control objective is to regulate all DG voltages to the reference value $1~pu$. 
A black cross mark represents a voltage violation on the DG whereas a black dot means the DG's voltage is within the normal range. Here we use $r$ to denote the average control reward of the last 5 control steps according to Eqn.~\ref{eqn:reward_fn}. All algorithms can regulate the voltages to the normal range in the small-size  microgrid-6 environment as shown in Fig. \ref{fig:voltage6}.

Fig. \ref{fig:voltage20} shows the performance comparison for the more challenging microgrid-20 case, due to dense network connections and complicated couplings among agents. It is thus not wise to control the voltage independently as IA2C does, which indeed leads to a bad performance. 
MPC performs well but it relies on an accurate model and it is very computationally expensive (19 times longer inference time than PowerNet) as it solves an online nonlinear optimization problem at each time step, requiring great computation power.
FPrint, DIAL, and CommNet also incur voltage violations, which shows that simply including neighbors' policies is not sufficient for a complicated test case as microgrid-20.
DIAL and CommNet's failure imply that simply summing all the communication information and computing the immediate reward according to undiscounted global reward can cause agents fail to learn effective communication protocols in a large-scale and cooperative microgrid environment. ConseNet fails to regulate the voltages and the controlled voltages are still more diverged compared to PowerNet. 
\vspace{-8pt}

\subsection{Robustness to load variations and agent mis-connections}

After 10 thousand episodes of training, we evaluate the trained policy for 20 times under different load disturbance with the same random seed for each agent in every episode, while sharing different test seeds for different episodes. Besides the state-of-the-art MARL algorithms, we also compare PowerNet with the conventional model-based method \cite{lou2016distributed} in terms of voltage control performance. The comparison results are summarized in Table \ref{tb:eval_return}, which shows the average episode return over 20 test episodes of the MARL algorithms and the conventional method (MPC).

It is clear that PowerNet consistently achieves the best execution performance over other MARL methods in both scenarios under different load variations. 
MPC method achieves good performance while it suffers from long convergence time (approximately 19.6 times longer than PowerNet). IA2C also does not perform well in both scenarios which is consistent with the training curves  shown in Fig.~\ref{fig:overall_performance}. Other state-of-the-art MARL benchmarks achieve similar, reasonable performance but our PowerNet outperforms them with a significant margin.

\begin{table*}[!th]
\renewcommand{\arraystretch}{1.4}
\centering
\caption{Performance comparison between trained MARL policies and the conventional model-based method under different load disturbance. The reward is the average reward over 20 evaluation episodes. Best values are bolded.}
\label{tb:eval_return}
\begin{tabular}{c|c|c|c|c|c|c|c|c|c}
\hline
Load Disturbance        & Network      & PowerNet & MPC \cite{lou2016distributed} &IA2C  & FPrint & ConseNet & CommNet & DIAL  & MADDPG \\ \hline
\multirow{2}{*}{10\%}  & Microgrid-6  & \textbf{0.240}    
& 0.141     & 0.206      & 0.236     & 0.219   & 0.223     & 0.222    & 0.221    \\ \cline{2-10} 
                      & Microgrid-20 & \textbf{0.781}   & 0.642    & 0.447      & 0.711     & 0.705       & 0.700     & 0.706   & 0.436     \\ \hline
\multirow{2}{*}{15\%} & Microgrid-6  & \textbf{0.238}             & 0.139    & 0.206    & 0.236        & 0.221       & 0.220     & 0.220   & 0.220     \\ \cline{2-10} 
                      & Microgrid-20 & \textbf{0.771}        & 0.632     & 0.474      & 0.702       & 0.692       & 0.690     & 0.697   & 0.422     \\ \hline
\multirow{2}{*}{25\%} & Microgrid-6  &  \textbf{0.236}    &  0.134 & 0.205 & 0.233  & 0.223    & 0.220   & 0.221 & 0.220       \\ \cline{2-10} 
                      & Microgrid-20 & \textbf{0.740}    &  0.598 & 0.458 & 0.660  & 0.645    & 0.649   & 0.663 & 0.438        \\ \hline
\end{tabular}
\end{table*}

\begin{figure*}[!ht]
  \centering
  \includegraphics[width=0.75\textwidth]{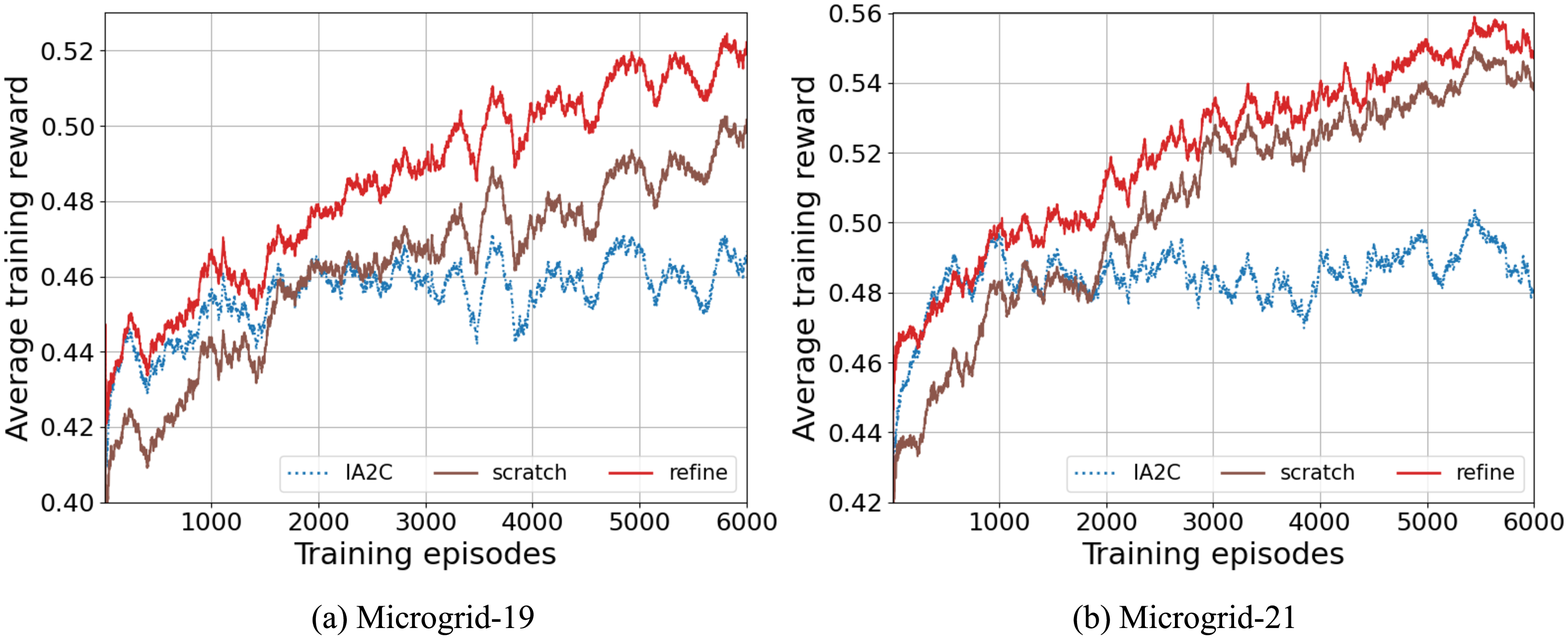}
  \caption{MARL training curves compared with training from scratch and adapting from a pre-trained microgrid-20 model for (a) microgrid-19 and (b) microgrid-21 systems. The lines show the average reward per training episode which are smoothed over the past 100 episodes.}
  \label{fig:plot_der21}
  \vspace{-5pt}
\end{figure*}

Furthermore, Fig.~\ref{fig:plot_der21} shows the comparison of the training curves between training from scratch and training from the pre-trained model when disconnecting one DG unit (microgrid-19) or adding a new DG unit (microgrid-21). The results show that the trained microgrid-20 can greatly facilitate the policy adaptations when a DG is disconnected or added to the microgrid. Since our algorithm is decentralized and on-policy, we do not need to re-train the whole network from scratch and we can just modify the involved DNN layers that are related to the topology change. For example, if a new DG unit is added, we just initialize and insert a new policy layer to the existing policy network and then link it to its neighboring layers. The other policy network will be loaded from the existing trained weights and the whole systems will be adapted.
From Fig.~\ref{fig:plot_der21}, we can see that the trained model facilitates the adaptation and achieve promising results in a shorter time with a better performance as compared to training from scratch.
\vspace{-8pt}

\subsection{Scalability to larger powergrid}
\begin{figure}[!ht]
  \centering
  \includegraphics[width=0.4\textwidth]{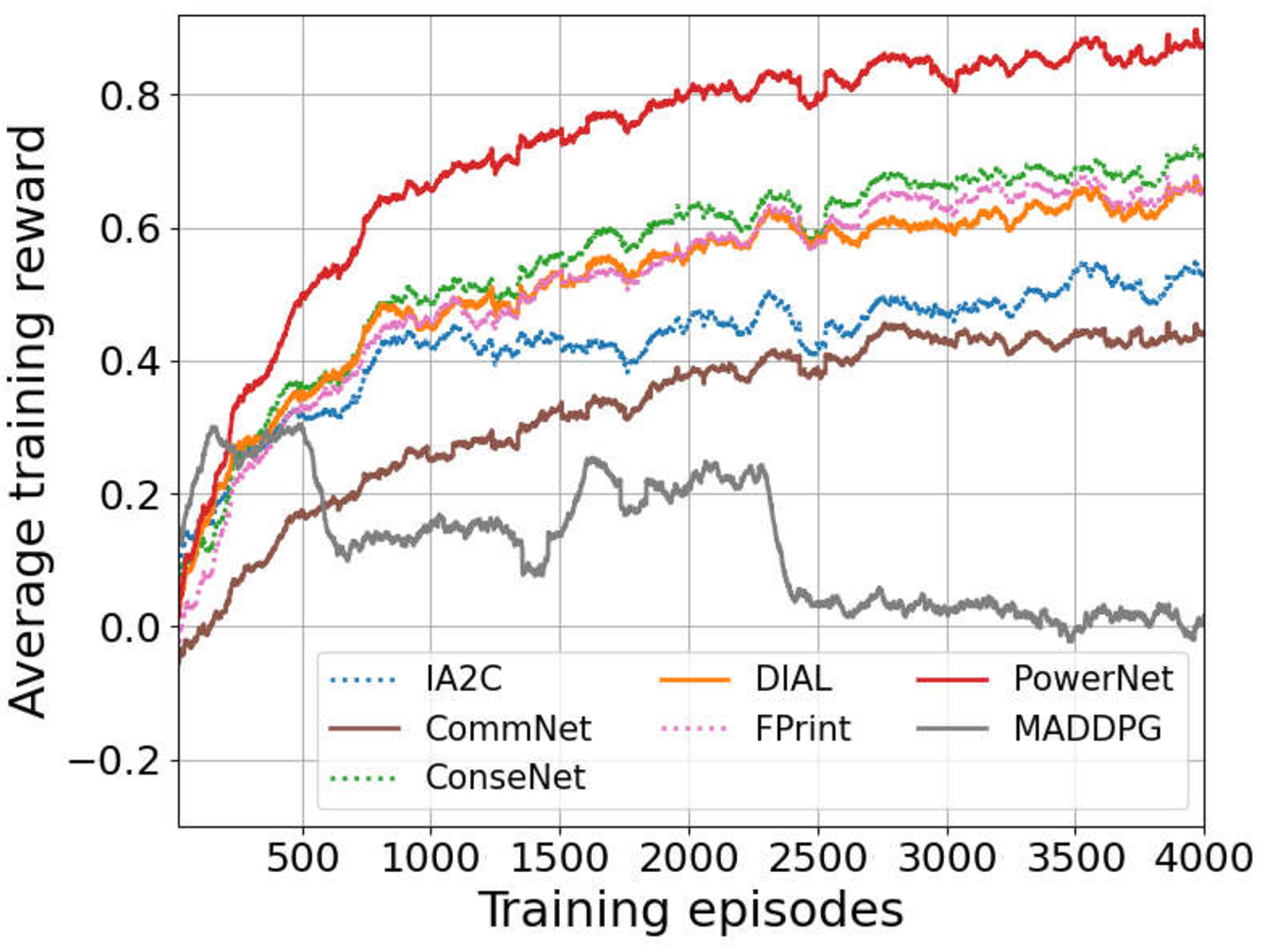}
  \caption{MARL training curves comparison between trained MARL policies for microgrid-40 systems. The lines show the average reward per training episode which are smoothed over the past 100 episodes. }
  \label{fig:plot_der40}
  \vspace{-15pt}
\end{figure}

In this subsection, we further investigate the scalability of PowerNet by increasing the number of DG units to 40.  As shown in  Fig.~\ref{fig:plot_der40}, PowerNet still works well and consistently outperforms other MARL benchmarks. It is interesting that CommNet fails to converge and becomes  worse than IA2C, which might be due to that it takes the mean of communication message instead of encoding them.

The inference time for an agent to generate an action is as shown in Table \ref{tb:infer_time} for microgrid-6, microgrid-20, and microgrid-40 networks, respectively. PowerNet takes only 35.8 ms to determine an action  in a large-scale microgrid with 40 agents, which is practically sound to meet real-time requirements. Since PowerNet is a decentralized algorithm, the ratio of inference time over network size is stable, confirming its great scalability.

\begin{table}[!ht]
\renewcommand{\arraystretch}{1.4}
\centering
\caption{Inference time and time/size ratio of PowerNet for different microgrid networks.}
\label{tb:infer_time}
\begin{tabular}{c|c|c|c}
\hline
                   & Microgrid-6 & Microgrid-20 & Microgrid-40 \\ \hline
Inference time     &    7.9$~ms$         &    16.8$~ms$          &    35.8$~ms$          \\ \hline
Time/size  &    1.32         &   0.84           &   0.90         \\ \hline
\end{tabular}
\vspace{-15pt}
\end{table}

\section{Conclusions} \label{sec:6}
In this paper, we formulated the secondary voltage control in inverter-based microgrid systems as a MARL problem. A novel on-policy, cooperative MARL algorithm, PowerNet, was developed by incorporating a differentiable, learning-based communication protocol, a spatial discount factor, and an action smoothing scheme.  Comprehensive experiments were conducted that show  the proposed PowerNet  outperforms other state-of-the-art approaches in terms of convergence speed and voltage control performance. In our future work, we will develop a more realistic simulation environment by incorporating  data from real-world power systems. We will also consider more extreme system disturbances and investigate safety-guaranteeing schemes to ensure safety during learning.
\vspace{-5pt}

\bibliography{ref.bib}
\bibliographystyle{IEEEtran}
\end{document}